\documentstyle[11pt,aaspp4]{article} 
\begin{document}
\title{Non-Radiative Accretion and Thermodynamics} 
\author{Andrei Gruzinov}
\affil{Physics Department, New York University, 4 Washington Place, New York, NY 10003}

\begin{abstract}

It has been suggested that the laws of thermodynamics are violated by what we have called a convection-dominated accretion flow (or a 1/2-law accretion flow) -- an accretion flow characterized by a constant outflow of energy. We show that both the 1/2-law flow and the Bondi flow (also known as ADAF, advection dominated accretion flow) are thermodynamically admissible.

\end{abstract}

\section{An unsolved problem: the rate of non-radiative quasi-spherical accretion}

We want to calculate the rate of quasi-spherical non-radiative accretion onto a black hole of Schwarzschild radius $R_S=2GM/c^2$. We assume that $R_S\ll R_A$, where the accretion radius $R_A=2GM/c_s^2$. Here $c_s$ is the speed of sound at infinity, where there is gas at rest, with mass density $\rho $. 

A standard answer is the Bondi (1952) rate: 
\begin{equation}
\dot{M} _{\rm Bondi}\sim R_A^2c_s\rho .
\end{equation}
This indeed is the correct rate for an ideal spherically symmetrical stationary flow of compressible fluid with adiabatic index less than 5/3. Under spherical symmetry, for an ideal gas, one can get Bondi in a direct numerical simulation.

But: (i) the spherical symmetry is almost certainly broken by turbulence, (ii) turbulence makes the flow non-stationary, (iii) magnetic fields are probably generated, (iv) dissipative processes appear. Then Bondi is not immediately applicable, but it can still be a good estimate. Indeed, the rate of mass accretion should be about $\dot{M}\sim R_S^2c\rho _S$, where $\rho _S$ is the density near the hole, where everything goes in at about the speed of light. Now we want to estimate $\rho _S$ in terms of $\rho $. This can be done by requiring that the mass flux be constant, $R^2v(R)\rho(R)=$const. Assuming that $v(R)$ is close to the free fall, $\propto R^{-1/2}$, we get $\rho(R) \propto R^{-3/2}$, and therefore the Bondi rate. 

Let us do it again: the rate of mass accretion should be about $\dot{M}\sim R_S^2c\rho _S$. To estimate $\rho _S$, assume that thermal energy generated by accretion is transported outward by a Shakura-Sunyaev-like turbulence (the flow is non-radiative) at a constant rate $\sim R^2c_s^3(R)\rho(R)=$const. This gives a 1/2-law density profile $\rho(R) \propto R^{-1/2}$, and a much smaller mass accretion rate 
\begin{equation}
\dot{M} _{\rm 1/2-law}\sim R_AR_Sc_s\rho .
\end{equation}
This small accretion rate, or equivalently a shallow density profile, is the main feature of CDAF, which is a particular worked out example of the 1/2-law flow (Gruzinov 2001, Quataert \& Gruzinov 1999, Balbus \& Hawley 2002 and references therein). I take CDAF to mean constant energy flux and not all of math details in all the papers (including ours, Quataert \& Gruzinov 1999) e.g., marginal stability, ....  Very reasonable to admit that math details are different in the magnetic case, but one can still have a constant energy flux as a defining feature of the flow.

Which answer is correct? The three dimensional time-dependent MHD (magnetohydrodynamics) simulation is the right way to solve the problem (Balbus \& Hawley 2002). Or is it?

To see what are the typical plasma parameters, consider accretion onto the black hole at the center of the Galaxy. A recent Chandra detection of quiescent Sgr A$^*$ (Baganoff et al 2001) can be understood as thermal bremsstrahlung coming from about the vicinity of accretion radius $R_A\sim 10^{17}cm$, with gas number density at accretion radius $n\sim 10^3{\rm cm}^{-3}$. (Quataert \& Gruzinov 2000, Quataert 2002). Then, assuming Bondi, we can estimate the gas temperature and density at smaller radii. Say at $R=100R_S=10^{14}{\rm cm}$, we have a plasma of temperature $T(R)\sim 1$MeV, and $n(R)\sim (R/R_A)^{-3/2}n\sim 3\times 10^7{\rm cm}^{-3}$. This gives a mean free path (from Coulomb collisions) $\lambda \sim 3\times 10^{16}{\rm cm}\gg R$. \footnote{ $\lambda \sim 10^{12}{\rm cm}(T/1{\rm eV})^2/(n/1{\rm cm}^{-3})$} The mean free path is even larger if the 1/2-law density profile is used. Thus we have a collisionless plasma. MHD is not applicable (because particles make large excursions along magnetic field lines).

A direct numerical simulation of a collisionless plasma seems impossible today -- even the MHD simulations have inadequate resolution to see if the Bondi, $n\propto R^{-3/2}$, or the 1/2-law, $n\propto R^{-1/2}$, flow forms (Hawley et al 2001). In this situation, it would be of great importance, if thermodynamic arguments could be used to estimate the rate of accretion (or equivalently the density profile). Indeed, Balbus and Hawley (2002) have argued that thermodynamics rules out CDAFs.

This cannot be correct because CDAFs were obtained in a thermodynamically consistent simulation. (A simulation is thermodynamically consistent if entropy is growing, that is ``all viscosities are positive''). Thus the 1/2 law is thermodynamically allowed. But this is not the point of the present note. 

Here we will show that all accretion rates smaller than Bondi but larger than the 1/2-law are thermodynamically admissible. Therefore, until a collisionless plasma simulation is done, one won't be able to say what is the true accretion rate in Sgr A$^*$ and in many other Dim Black Holes.

\section{Thermodynamically admissible collisionless non-radiative quasi-spherical accretion flows}

{\bf Bondi}: We start with the standard lore. In a collisionless plasma, electromagnetic field will appear. Thermodynamics does not contradict the assumption that a weak small-scale magnetic field is the only field generated. This field will make an effective mean free path small. Plasma becomes a fluid. Thermodynamics does not contradict the assumption that the resulting flow will be spherical, stationary, with negligible dissipation. Under these assumptions, we get Bondi, equation (1). 

{\bf 1/2 law}: Thermodynamics does not contradict the assumption that no electromagnetic fields are generated. Then the plasma is just a collection of non-interacting particles in a gravitational potential. At large times $\sim t$, particles at distances $R\sim c_st$ will have a chance of flying into the hole. To hit the hole, the particle's velocity should lie within the loss cone of opening angle $\theta \sim \sqrt{R_AR_S}/R$. The total mass accreted into the hole at times $\sim t$ is $\sim \rho R^3\theta ^2$, and the corresponding mass accretion rate is the 1/2-law rate given by equation (2).

Clearly, by adjusting our (thermodynamically admissible) assumptions about the nature of the flow, we can get any accretion rate between Bondi and 1/2-law.

\section{Conclusions}

\begin{itemize}
\item Thermodynamics alone cannot solve the problem of  non-radiative quasi-spherical accretion flows. Very different flow regimes, Bondi, 1/2-law, and intermediate (Gruzinov 1999) are thermodynamically admissible. This is true for fluids as well as for collisionless plasmas.

\item Three dimensional MHD numerical simulations is the most reliable approach today. If adequate resolution is achieved (to reach the self-similar flow regime which almost certainly exists), the resulting rate of mass accretion will have to be taken seriously. 

\item However, in many astronomically interesting cases the plasma is collisionless, and even 3D MHD is potentially misleading. Large non-MHD terms (say Shakura-Sunayev-like transport coefficients included into the basic model) might be present. The only way to get a true rate of mass accretion in Sgr A$^*$ and in other Dim Black Holes is direct simulation of collisionless plasma, which today looks like sheer impossibility.

\end{itemize}

\acknowledgements

I thank Eliot Quataert for useful discussions; several sentences in this note are his. I thank my NYU colleagues Roman Scoccimarro and Matias Zaldarriaga for bringing the preprint of Balbus and Hawley to my attention and demanding explanations.

\end{document}